\documentclass[journal]{IEEEtran}

\ifCLASSINFOpdf
\else
\fi
\usepackage{amssymb}  
\usepackage{graphics} 
\usepackage{epsfig} 
\usepackage{mathptmx} 
\usepackage{times} 
\usepackage{amsmath} 

\usepackage{floatrow}
\usepackage{multirow}

\usepackage{xcolor}
\DeclareMathAlphabet{\mathcal}{OMS}{cmsy}{b}{n}
\DeclareMathAlphabet{\mathcal}{OMS}{cmsy}{m}{n}
\newtheorem{theorem}{\indent Theorem}
\newtheorem{lemma}{\indent Lemma}
\newtheorem{corollary}{\indent Corollary}
\newtheorem{proposition}{\indent Proposition}

\newtheorem{example}{\indent Example}
\newtheorem{remark}{\indent Remark}
\newtheorem{problem}{\indent Problem}
\newtheorem{algorithm}{\indent Algorithm}
\newtheorem{assumption}{\indent Assumption}
\begin{document}

\title{A Quantum Hamiltonian Identification Algorithm: Computational Complexity and Error Analysis}
%
%
%

\author{Yuanlong~Wang,
~Daoyi~Dong, 
Bo~Qi,
Jun~Zhang, 
Ian~R.~Petersen, 
Hidehiro~Yonezawa

\thanks{This work was supported by the Australian Research Council's Discovery Projects funding scheme under Project DP130101658, Laureate Fellowship FL110100020, Centres of Excellence CE110001027 and the National Natural Science Foundation of China under Grant No. 61374092.}
\thanks{Y. Wang and H. Yonezawa are with the School of Engineering and Information Technology, University of New South Wales, Canberra, ACT 2600, Australia, and also with the Centre for Quantum Computation and Communication Technology, Australian Research Council, Canberra, ACT 2600, Australia (e-mail: yuanlong.wang.qc@gmail.com; h.yonezawa@adfa.edu.au).}%
\thanks{D. Dong and I. R. Petersen are with the School of Engineering and Information Technology, University of New South Wales, Canberra, ACT 2600, Australia (e-mail: daoyidong@gmail.com; i.r.petersen@gmail.com).} 
\thanks{B. Qi is with the Key Laboratory of Systems and Control, Academy of Mathematics and Systems Science, Chinese Academy of Sciences, Beijing
100190, China and with the University of Chinese Academy of Sciences,
Beijing 100049, China (e-mail: qibo@amss.ac.cn).}
\thanks{J. Zhang is with the Joint Institute of UM-SJTU, Shanghai Jiao Tong University, Shanghai 200240, China
(e-mail: zhangjun12@sjtu.edu.cn).}
}

\maketitle

\begin{abstract}
Quantum Hamiltonian identification is important for characterizing the dynamics of quantum systems, calibrating quantum devices and achieving precise quantum control. In this paper, an effective two-step optimization (TSO) quantum Hamiltonian identification algorithm is developed within the framework of quantum process tomography. In the identification method, different probe states are inputted into quantum systems and the output states are estimated using the quantum state tomography protocol via linear regression estimation. The time-independent system Hamiltonian is reconstructed based on the experimental data for the output states. The Hamiltonian identification method has computational complexity $O(d^6)$ where $d$ is the dimension of the system Hamiltonian. An error upper bound $O(\frac{d^3}{\sqrt{N}})$ is also established, where $N$ is the resource number for the tomography of each output state, and several numerical examples demonstrate the effectiveness of the proposed TSO Hamiltonian identification method.
\end{abstract}

\begin{IEEEkeywords}
Quantum system, Hamiltonian identification, process tomography, computational complexity.
\end{IEEEkeywords}

%
\IEEEpeerreviewmaketitle

\section{INTRODUCTION}

\IEEEPARstart{A}{s} quantum technology develops, there is an increasing demand for characterizing an unknown quantum process since it is vital to verify and benchmark quantum devices for quantum computation, communication and metrology \cite{Nielsen and Chuang 2000}. The standard solution to characterizing a quantum process is Quantum Process Tomography (QPT), wherein usually known input quantum states (probe states) are applied to the process and the output states are measured to reconstruct the quantum process \cite{jf and hradil 2001}-\cite{add0}. Hamiltonian identification for closed quantum systems is a special class of QPT that corresponds to a unitary quantum process and is an essential component to characterize the dynamics of a quantum system.

System identification has been widely investigated in classical (non-quantum) systems and control theory, and many identification algorithms have been developed to estimate unknown dynamical parameters of linear or nonlinear input-output systems \cite{chen 2014}-\cite{sontag 2009}. In recent years, the problem of quantum system identification has attracted more and more attention due to the rapid development of emerging quantum technology \cite{hidehiro 2012,xiang 2011} and increasing demand of characterizing quantum devices. For example, a framework for quantum system identification has been established in \cite{burgarth 2012} to classify how much knowledge about a quantum system is attainable from a given experimental setup. Gu\c{t}\u{a} and Yamamoto \cite{guta 2016} considered a class of passive linear quantum input-output systems, and investigated the problem of identifiability and how to optimize the identification precision by preparing good input states and performing appropriate measurements on the output states.

In this paper, we focus on the problem of quantum Hamiltonian identification (QHI), which is a key task in characterizing the dynamics of quantum systems and achieving high-precision quantum control. There exist some results on QHI and various aspects of QHI have been investigated \cite{add1}-\cite{sone 2016}. For example, a symmetry-preserving observer has been developed for the Hamiltonian identification of a two-level quantum system \cite{bonnabel 2009}. The identifiability problem for a Hamiltonian corresponding to a dipole moment has been investigated \cite{bris 2007} and the question of how to utilize quantum control to identify such Hamiltonian has been addressed \cite{leghtas 2012}. Closed-loop learning control has been presented to optimally identifying Hamiltonian information \cite{geremia 2002} and compressed sensing has been proposed to enhance the efficiency of identification algorithms for Hamiltonian with special structures \cite{shabani 2011,rudinger 2015}. Several Hamiltonian identification algorithms have been developed using only measurement in a single fixed basis \cite{schirmer 2009}-\cite{sergeevich 2011}. Wang \emph{et al.} \cite{wang 2015} utilized dynamical decoupling to identify Hamiltonians for quantum many-body systems with arbitrary couplings. Cole \emph{et al.} \cite{cole 2005} discussed the estimation error in identifying a two-state Hamiltonian  and Zhang \emph{et al.} \cite{zhang 2014} presented a QHI protocol using measurement time traces. Most of these existing results have limitations for practical applications (e.g., estimating a single parameter \cite{sergeevich 2011,yuan 2015}, identifying special Hamiltonian \cite{schirmer 2009,cole 2005}), and there are few theoretical results on the analysis of computational complexity and upper bounds on estimation errors. This paper presents an identification algorithm for general time-independent Hamiltonians and analyzes its computational complexity and upper bounds on estimation errors.

Our quantum Hamiltonian identification is presented within the framework of quantum process tomography. Some different input states are prepared for quantum systems and the corresponding output states are measured after a fixed time evolution under the Hamiltonian to be identified. These output states are reconstructed using the quantum state tomography technique via linear regression estimation (LRE) \cite{qi 2013}. Using the information of estimated output states, the Hamiltonian is reconstructed via an identification algorithm. The main contributions of this paper are summarized as follows.

\begin{itemize}
\item The quantum Hamiltonian identification (QHI) problem is formulated within the framework of quantum process tomography (QPT) and several relevant points in QPT are clarified in order to present an efficient QHI algorithm.
\item A Two-Step Optimization (TSO) identification algorithm is presented and its computational complexity is analyzed. Our identification algorithm has the computational complexity $O(d^6)$ where $d$ is the dimension of the quantum system.
\item Analytical results of estimation error are presented and an error upper bound is established as $O(\frac{d^3}{\sqrt{N}})$, where $N$ is the resource number in the tomography of each output state.
\item Numerical examples are presented to demonstrate the performance of our QHI algorithm. It is then compared with the QHI method using measurement time traces in \cite{zhang 2014}, and our identification algorithm shows an efficiency advantage over the method in \cite{zhang 2014} in terms of the computational time.
\end{itemize}

The structure of this paper is as follows. In Section \ref{Sec2} we present some preliminaries and briefly introduce QPT. Section \ref{Sec3} formulates the QHI problem within the framework of QPT. Section \ref{Sec4} presents a TSO Hamiltonian identification algorithm and analyzes the computational complexity. Section \ref{err} analyzes the estimation error theoretically and establishes an upper bound. In Section \ref{secsim}, we present two numerical examples to demonstrate performance and also compare our identification algorithm with the QHI method using time traces in \cite{zhang 2014}. Section \ref{secfinal} concludes this paper.

Notation:
$a^*$ denotes the conjugate of $a$; $A_{m\times n}$ denotes an $m$-row and $n$-column matrix; $A^T$ is the transpose of $A$; $A^\dagger$ is the conjugate and transpose of $A$; $\mathbb{H}$ denotes a Hilbert space; $\mathbb{R}$ and $\mathbb{C}$ are the sets of all real and complex numbers, respectively; $I$ is the identity matrix (dimension omitted if without ambiguity);
$||A||$ denotes the Frobenius norm of $A$; $\text{Tr}(A)$ is the trace of $A$; $|\psi\rangle$ is a unit complex vector representing a quantum (pure) state; $\rho$ is a density matrix representing a quantum state; $\hat a$ is
the estimate of $a$; $\langle A, B\rangle$ represents the inner product of $A$ and $B$ defined as $\langle A, B\rangle=\text{Tr}(A^\dagger B)$; $\langle a, b\rangle$ denotes the inner product of $a$ and $b$ with $\langle a,
b\rangle=a^\dagger b$; $\mathbb{C}_d$ is the set of all $d$-dimension complex vectors; $\mathbb{C}_{d\times d}$ is the set of all $d\times d$ complex matrices; $\text{vec}(\cdot)$ denotes the vectorization function; $\text{vec}^{-1}
(\cdot)$ is the inverse function of vectorization from $\mathbb{C}_{d^2}$ to $\mathbb{C}_{d\times d}$; $A\otimes B$ denotes the tensor product of $A$ and $B$; $\text{Tr}_1(X)$ means partial trace on space $\mathbb{H}_1$ where
$X\in\mathbb{H}_1\otimes\mathbb{H}_2$; $\delta$ is the Dirac Delta function; $i$ as a subscript means an integer index, otherwise $i$ means imaginary unit; i.e., $i=\sqrt{-1}$.

\section{PRELIMINARIES AND QUANTUM PROCESS TOMOGRAPHY}\label{Sec2}
\subsection{Matrix and Vectorization Fundamentals}
For a matrix $A_{m\times n}=[a_{ij}]$, its Frobenius norm is defined as $$||A||=\sqrt{\sum_{i=1}^m\sum_{j=1}^na_{ij}^2}= \sqrt{\text{Tr}(A^\dagger A)}.$$ Two important properties of the Frobenius norm are:
\begin{equation}\label{propertyhs1}
||A||=||UA||,
\end{equation}
\begin{equation}\label{propertyhs2}
||AB||\leq||A||\cdot||B||,
\end{equation}
where $U$ is any $m\times m$ unitary matrix.

The tensor product of matrices $A_{m\times n}=[a_{ij}]$ and $B_{p\times q}=[b_{kl}]$ is defined as follows:
\begin{equation}
A\otimes B=\left(\begin{array}{*{4}{c}}
a_{11}B & a_{12}B & \cdots & a_{1n}B \\
a_{21}B & a_{22}B & \cdots & a_{2n}B \\
\multicolumn{4}{c}{\dotfill} \\
a_{m1}B & a_{m2}B & \cdots & a_{mn}B \\
\end{array}\right)_{mp\times nq}.
\end{equation}
We introduce the vectorization function $\text{vec}: \mathbb{C}_{m\times n}\mapsto \mathbb{C}_{mn}$. For a matrix $A_{m\times n}=[a_{ij}]$, $$\text{vec}(A_{m\times n})=[a_{11},a_{21},...,a_{m1},a_{12},...,a_{m2},...,a_{1n},...,a_{mn}]^T.$$ The function $\text{vec}(\cdot)$ (also denoted as $|\cdot \rangle\rangle$ or $|\cdot)$ in the physics community) is linear. Its common properties are listed as follows \cite{horn 1985,watrous}:
\begin{equation}\label{property6}
\text{vec}(|a\rangle\langle b|)=|b\rangle^*\otimes|a\rangle,
\end{equation}
\begin{equation}\label{property2}
\text{vec}(AXB)=(B^T\otimes A)\text{vec}(X),
\end{equation}
\begin{equation}\label{property3}
\langle A, B\rangle=\langle \text{vec}(A), \text{vec}(B)\rangle,
\end{equation}
\begin{equation}\label{property4}
\text{Tr}_1(\text{vec}(A)\text{vec}(B)^{\dagger})=AB^{\dagger}.
\end{equation}
\begin{equation}\label{property5}
\text{Tr}_2(\text{vec}(A)\text{vec}(B)^{\dagger})=(B^{\dagger}A)^T.
\end{equation}
In this paper, we also define that $\text{vec}^{-1}(\cdot)$ maps a $d^2\times 1$ vector into a $d\times d$ square matrix rather than matrices with other sizes. In (\ref{property4}) and (\ref{property5}), $\text{Tr}_1(X)$ means partial trace on the space $\mathbb H_1$ where $X$ belongs to the space $\mathbb H_1\otimes \mathbb H_2$. Similarly $\text{Tr}_2(X)$ means partial trace on the space $\mathbb H_2$.

\subsection{Quantum System and Evolution}
The state of a closed quantum system can be described by a unit complex vector $|\psi\rangle$ in the underlying Hilbert space and its dynamics is governed by the Schr\"{o}dinger equation
\begin{equation}
i\frac{\partial}{\partial t}|\psi(t)\rangle=H|\psi(t)\rangle,
\end{equation}
where $H$ is the system Hamiltonian and we set $\hbar=1$ using atomic units in this paper. When the quantum system under consideration is an open quantum system or the quantum state is a mixed state, we need to use a Hermitian positive semidefinite matrix $\rho$ satisfying $\text{Tr}(\rho)=1$ to describe the quantum state. For a closed quantum system with state $|\psi\rangle$, we have $\rho=|\psi\rangle\langle\psi|$. Its evolution from the initial state $\rho(0)$ to $\rho(t)$ at time $t$ can be determined by a unitary propagator $U$:
\begin{equation}\label{eq61}
\rho(t)=U(t)\rho(0)U^\dagger(t),
\end{equation}
where $U(t)=e^{-iHt}$ if $H$ is independent of $t$.

For an open quantum system, the dynamics of its state can be described by a master equation. Alternatively, the transformation from an input state $\rho_{in}$ to an output state $\rho_{out}$ is given by Kraus operator-sum representation \cite{Nielsen and Chuang 2000}
\begin{equation}\label{kraus}
\rho_{out}=\mathcal{E}(\rho_{in})=\sum_{i}A_i\rho_{in} A_i^\dagger,
\end{equation}
where the quantum operation $\mathcal{E}$ maps $\rho_{in}$ to $\rho_{out}$ and $\{A_i\}$ is a set of mappings from the input Hilbert space to the output Hilbert space with $\sum_i A_i^\dagger A_i\leq I$. In this paper, we only consider trace-preserving operations which means that the completeness relation
\begin{equation}\label{trpreserve}
\sum_i A_i^\dagger A_i= I
\end{equation}
is satisfied. In particular, we consider $d$-dimensional quantum systems and have $A_i\in\mathbb{C}_{d\times d}$.

\subsection{Quantum Measurement and Quantum State Tomography}
We aim to identify the system Hamiltonian $H$ from the input states (usually known) and the output states. To extract information from the output quantum states, a positive-operator valued measurement (POVM) is usually performed on these states. A POVM is a set $\{M_i\}$, where all the elements are Hermitian positive semidefinite and $\sum_iM_i=I$. When a set of POVM is performed, the probability of outcome $i$ occurring is determined by the Born Rule $p_i=\text{Tr}(\rho M_i)$. A special class of POVM are the projective measurement operators $\{P_i\}$, which are projectors satisfying $P_iP_j=\delta_{ij}P_j$.

In real experiments, it is impossible to implement infinitely many measurements. Hence, $p_i$ can only be approximated within a limited accuracy. The methodology for designing the measurement operators $\{M_i\}$ and estimating $\rho$ from experimental data is called quantum state tomography, where usually a large number of identified independent copies of $\rho$ are used. Common quantum state tomography methods include Maximum likelihood estimation \cite{paris 2004}-\cite{teo 2011}, Bayesian mean estimation \cite{paris 2004, bayesianslow} and linear regression estimation (LRE) \cite{qi 2013}. In this paper, the LRE method will be used in numerical simulation and error analysis although our QHI method is also applicable to other quantum state tomography methods. In the LRE method, the quantum state reconstruction problem is converted into a parameter estimation problem for a linear regression model and the least-squares method can be used to obtain estimates of the unknown parameters. The LRE method of quantum state tomography was first presented in \cite{qi 2013} and it has also been used to experimentally reconstruct quantum states for various tasks \cite{qi 2015,hou 2016}. Its advantages of high efficiency and an analytical error upper bound make it especially beneficial in presenting numerical results and error analysis for our TSO QHI method.

\subsection{Standard Quantum Process Tomography}
We rephrase the framework of general quantum process tomography in \cite{Nielsen and Chuang 2000} in the matrix form and later we will consider QHI problem under this framework.

By expanding $\{A_i\}$ in (\ref{kraus}) in a fixed family of basis matrices $\{E_i\}$, we obtain
\begin{equation}\label{actrans}
A_i=\sum_j c_{ij}E_j,
\end{equation}
and then $$\mathcal{E}(\rho)=\sum_{jk}E_j\rho E_k^\dagger x_{jk},$$ with $x_{jk}=\sum_i c_{ij}c_{ik}^*$. If we define the matrix $C=[c_{ij}]$ and the matrix $X=[x_{ij}]$, then
\begin{equation}\label{xctrans}
X=C^TC^*,
\end{equation}
which indicates that $X$ must be Hermitian and positive semidefinite. $X$ is called the {\itshape process matrix} \cite{brien etal 2004}. The completeness constraint equation (\ref{trpreserve}) becomes
\begin{equation}\label{ptrace1}
\sum_{j,k}x_{jk}E_k^{\dagger}E_j=I.
\end{equation}
It is difficult to further simplify this relationship before the structure of $\{E_i\}$ is determined. Note that the matrix $X$ and the process $\mathcal{E}$ are in a one-to-one correspondence. Hence, we can obtain a full characterization of $\mathcal{E}$ by reconstructing $X$ \cite{Nielsen and Chuang 2000}. 

Let $\{\rho_m\}$ be a complete basis set of $\mathbb{C}_{d\times d}$. For example, all Pauli matrices $\sigma_{x}=\left(\begin{smallmatrix}
  0 & 1  \\
  1 & 0  \\
\end{smallmatrix}\right)$, $\sigma_{y}=\left(\begin{smallmatrix}
  0 & -i  \\
  i & 0  \\
\end{smallmatrix}\right)$ and $\sigma_{z}=\left(\begin{smallmatrix}
  1 & 0  \\
  0 & -1  \\
\end{smallmatrix}\right)$, together with $I_{2\times 2}$, form a complete basis set of $\mathbb{C}_{2\times 2}$. If we let $\{\rho_m\}$ be linearly independent matrices (with respect to addition between matrices, and multiplication between a scalar and a matrix) and we input $\rho_m$ to the process, then each process output can be expanded uniquely in the basis set $\{\rho_n\}$; i.e.,
\begin{equation}\label{eqlambda}
\rho_{out}=\mathcal{E}(\rho_{in})=\mathcal{E}(\rho_m)=\sum_n \lambda_{mn}\rho_n.
\end{equation}
For simplicity, we choose $\{\rho_n\}$ to be the same set as $\{\rho_m\}$ although they could be different. We then need to find the relationship between $X$ and $\lambda$, which is independent of the bases $\{E_i\}$. Considering the effects of the bases $\{\rho_n\}$ on $\{\rho_m\}$, we have
\begin{equation}\label{betadef}
E_j \rho_m E_k^\dagger=\sum_n\beta_{mn}^{jk}\rho_n.
\end{equation}
Hence, $$\sum_n\sum_{jk}\beta_{mn}^{jk}\rho_n x_{jk}=\sum_n \lambda_{mn}\rho_n.$$ From the linear independence of $\{\rho_n\}$, one can obtain
\begin{equation}
\sum_{jk}\beta_{mn}^{jk} x_{jk}=\lambda_{mn}.
\end{equation}
To rewrite this equation into a compact form, define the matrix $\Lambda=[\lambda_{mn}]$ and arrange the elements $\beta_{mn}^{jk}$ into a matrix $B$:
\begin{equation}\label{matrixB}
B=\left(\begin{array}{*{7}{c}}
\beta_{11}^{11} & \beta_{11}^{21} & \cdots & \beta_{11}^{12} & \beta_{11}^{22} & \cdots & \beta_{11}^{d^2d^2} \\
\beta_{21}^{11} & \beta_{21}^{21} & \cdots & \beta_{21}^{12} & \beta_{21}^{22} & \cdots & \beta_{21}^{d^2d^2} \\
\multicolumn{7}{c}{\dotfill} \\
\beta_{12}^{11} & \beta_{12}^{21} & \cdots & \beta_{12}^{12} & \beta_{12}^{22} & \cdots & \beta_{12}^{d^2d^2} \\
\beta_{22}^{11} & \beta_{22}^{21} & \cdots & \beta_{22}^{12} & \beta_{22}^{22} & \cdots & \beta_{22}^{d^2d^2} \\
\multicolumn{7}{c}{\dotfill} \\
\beta_{d^2d^2}^{11} & \beta_{d^2d^2}^{21} & \cdots & \beta_{d^2d^2}^{12} & \beta_{d^2d^2}^{22} & \cdots & \beta_{d^2d^2}^{d^2d^2} \\
\end{array}\right)_{d^4\times d^4}
\end{equation}
so that we have
\begin{equation}\label{eq3}
B\text{vec}(X)=\text{vec}(\Lambda).
\end{equation}
Here, $B$ is determined once the bases $\{E_i\}$ and $\{\rho_m\}$ are chosen, and $\Lambda$ is obtained from experimental data. $B$, $X$ and $\Lambda$ are in general complex matrices. Note that $X$ should be Hermitian and positive semidefinite and satisfy the constraint (\ref{ptrace1}). Hence, direct inversion or pseudo-inversion of $B$ may fail to generate a physical solution. We try to find a physical estimate $\hat X$ which will generate an output $\hat\rho'_{ge}$ as close as possible to the estimated results $\hat\rho'$ from quantum state tomography. Because $\hat\rho'_{ge}$ and $\hat\rho'$ are characterized by $\hat\Lambda_{ge}$ and $\hat\Lambda$ separately, we should minimize $||\hat\Lambda_{ge}-\hat\Lambda||$. Since $$||\hat\Lambda_{ge}-\hat\Lambda||=||\text{vec}(\hat\Lambda_{ge})-\text{vec}(\hat\Lambda)||=||B\text{vec}(\hat X)-\text{vec}(\hat\Lambda)||,$$ we will take $||B\text{vec}(\hat X)-\text{vec}(\hat\Lambda)||$ as a performance index.

The problem is now the following optimization problem:
\begin{problem}\label{problem1}
Given the matrix $B$ and experimental data $\hat\Lambda$, find a Hermitian and positive semidefinite estimate $\hat{X}$ minimizing $||B\text{vec}(\hat X)-\text{vec}(\hat\Lambda)||$, such that (\ref{ptrace1}) is satisfied.
\end{problem}

It is difficult to obtain an analytical solution to Problem \ref{problem1}. In this paper, we do not directly solve Problem \ref{problem1} since the problem of QHI can be further specified based on Problem \ref{problem1}. After one obtains an estimate $\hat X$, it is straightforward to obtain Kraus operators $\{\hat A_i\}$. Since $\hat X$ is Hermitian, it has spectral decomposition $$\hat X=\sum_{i=1}^{d^2}u_i|v_i\rangle\langle v_i|,$$ where $u_i$ are real eigenvalues. Then $$\hat C^T=\sum_{i=1}^{d^2}\sqrt{u_i}|v_i\rangle\langle v_i|,$$ and $$\hat A_i=\sum_j c_{ij}E_j.$$

Though $X$ and $\mathcal{E}$ are in one-to-one correspondence, the notable property of the Kraus operator-sum representation is its non-uniqueness; i.e., there may be more than one different sets of Kraus operators that give rise to the same process $\mathcal{E}$. This comes from the procedure of decomposing $X$ into $C^TC^*$, which is in fact non-unique because $$X=C^TC^*=(C^TU^T)(U^*C^*)$$ holds for any unitary $U$. Hence, the deduction of $C$ from $X$ is non-unique.

\section{PROBLEM FORMULATION OF HAMILTONIAN IDENTIFICATION}\label{Sec3}
The objective of this paper is to develop a new algorithm to identify a time-independent Hamiltonian $H$. If we compare (\ref{eq61}) with the Kraus representation (\ref{kraus}), it is clear that the unitary propagator $U(t)$ is the only Kraus operator. Then from (\ref{actrans}) we know that the matrix $C$ is a row vector. Hence, from (\ref{xctrans}) we know $X$ is of rank one. It is worth mentioning that, for any given process $\mathcal{E}$, although the Kraus operator-sum representation is not unique, the process matrix $X$ is in fact uniquely determined. Although there might be other Kraus operator-sum representations where the number of operators is more than 1, the conclusion that $X$ is of rank one is always true. When $X$ is of rank one, the semidefinite requirement is naturally satisfied. Let $X=gg^{\dagger}$ and $g=\text{vec}(G)$.

Now we need to determine basis sets $\{E_i\}$ and $\{\rho_m\}$. Proper choice of these basis sets can greatly simplify the QHI problem, and we thus choose both of them as the natural basis $\{|j\rangle\langle k|\}_{1\leq j,k\leq d}$, because the natural basis can simplify the completeness requirement (\ref{ptrace1}) and Problem \ref{problem1}. These advantages can be demonstrated as follows.

\begin{proposition}\label{ptracepropo}
If $\{E_i\}$ is chosen as the natural basis and the relationship between $i$, $j$ and $k$ is $i=(j-1)d+k$, then the completeness constraint reads $\text{Tr}_1X=I_d$.
\end{proposition}

The proof of Proposition \ref{ptracepropo} is presented in Appendix \ref{app0}.

The natural basis is also useful in transforming Problem \ref{problem1} into an optimization problem in a more convenient form:
\begin{problem}\label{problem2}
Given the matrix $B$ and experimental data $\hat\Lambda$, find a Hermitian and positive semidefinite estimate $\hat{X}$ minimizing $||\hat X-\text{vec}^{-1}(B^{-1}\text{vec}(\hat\Lambda))||$, such that constraint (\ref{ptrace1}) is satisfied.
\end{problem}

Problem \ref{problem2} is not necessarily equivalent to Problem \ref{problem1}. We need to determine when $B$ is invertible and when these two problems are equivalent. To answer these two questions, we give the following conditions to characterize $B$.
\begin{theorem}\label{binvert}
Let $\{E_i\}_{i=1}^{d^2}$ be a set of matrices in the space $\mathbb C_{d\times d}$ and let $\{\rho_m\}_{m=1}^{d^2}$ be a set of linearly independent bases of $\mathbb C_{d\times d}$. Define $B$ through (\ref{betadef}) and (\ref{matrixB}). Then $\{E_i\}$ is a set of linearly independent bases of $\mathbb C_{d\times d}$ if and only if $B$ is invertible.
\end{theorem}

\begin{theorem}\label{proequiv}
Let $\{E_i\}_{i=1}^{d^2}$ be a set of matrices in $\mathbb C_{d\times d}$ and let $\{\rho_m\}_{m=1}^{d^2}$ be a set of normal orthogonal bases of $\mathbb C_{d\times d}$. Define $B$ through (\ref{betadef}) and (\ref{matrixB}). Then $\{E_i\}$ forms normal orthogonal basis of $\mathbb C_{d\times d}$ if and only if $B$ is unitary.
\end{theorem}

The detailed proofs of Theorem \ref{binvert} and Theorem \ref{proequiv} are presented in Appendix \ref{app2} and Appendix \ref{app3}, respectively. Under the conditions in Theorem \ref{proequiv}, $B$ is unitary, and we have
\begin{equation*}
\begin{array}{cc}
||B\text{vec}(\hat X)-\text{vec}(\hat\Lambda)||&=||\text{vec}(\hat X)-B^{-1}\text{vec}(\hat\Lambda)|| \ \ \ \ \\
&=||\hat X-\text{vec}^{-1}(B^{-1}\text{vec}(\hat\Lambda))||,\\
\end{array}
\end{equation*}
which means Problem \ref{problem1} is equivalent to Problem \ref{problem2} in this case. The natural basis set satisfies the requirements in Theorem \ref{binvert} and Theorem \ref{proequiv}.

With the natural basis $\{|j\rangle\langle k|\}_{1\leq j,k\leq d}$ for $\{E_i\}$ and $\{\rho_m\}$, we have $$\text{Tr}_1(\text{vec}(G)\text{vec}(G)^{\dagger})=I_d=GG^{\dagger},$$ which means the completeness constraint (\ref{trpreserve}) is equivalent to the requirement that $G$ is unitary. Hence, we can transform Problem \ref{problem2} into the following problem which is critical for QHI.

\begin{problem}\label{problem3}
Assume that $\{\rho_m\}_{m=1}^{d^2}$ is a set of normal orthogonal bases of the space $\mathbb C_{d\times d}$, $\{E_i\}$ is chosen as $\{|j\rangle\langle k|\}_{1\leq j,k\leq d}$, and the relationship between $i$, $j$ and $k$ is $i=(j-1)d+k$. Given the unitary matrix $B$ and experimental data $\hat\Lambda$, find a unitary matrix $\hat G$ minimizing $||\text{vec}(\hat G)\text{vec}(\hat G)^{\dagger}-\text{vec}^{-1}(B^{\dagger}\text{vec}(\hat\Lambda))||$.
\end{problem}

\begin{remark}
Note that we can experimentally measure only Hermitian physical variables. Hence, we cannot directly use $|j\rangle\langle k|$ ($j\neq k$) as probe states. According to \cite{Nielsen and Chuang 2000}, when $j\neq k$, one can take $|j\rangle\langle j|$, $|k\rangle\langle k|$, $|+\rangle\langle +|$ and $|-\rangle\langle -|$ as inputs where $|+\rangle=(|j\rangle+|k\rangle)/\sqrt2$ and $|-\rangle=(|j\rangle+i|k\rangle)/\sqrt2$. Then $\mathcal{E}(|j\rangle\langle k|)$ can be obtained from
\begin{equation}\label{eq10}
\begin{array}{rl}
\mathcal{E}(|j\rangle\langle k|)=&\mathcal{E}(|+\rangle\langle +|)+i\mathcal{E}(|-\rangle\langle -|)\\
&\ -\frac{1+i}{2}\mathcal{E}(|j\rangle\langle j|)-\frac{1+i}{2}\mathcal{E}(|k\rangle\langle k|).\\
\end{array}
\end{equation}
\end{remark}

\section{HAMILTONIAN IDENTIFICATION ALGORITHM AND COMPUTATIONAL COMPLEXITY}\label{Sec4}
\subsection{Solution to Problem \ref{problem3}: Two-step Optimization (TSO)}\label{subsec2}

The direct solution to Problem \ref{problem3} is difficult \cite{wang 2016} and we split it into two sub-problems (which is the reason we name our method Two-Step Optimization):
\addtocounter{problem}{-1}
\renewcommand{\theproblem}{\arabic{problem}$.1$}
\begin{problem}\label{subproblem1}
Let $\hat D=\text{vec}^{-1}(B^\dagger \text{vec}(\hat\Lambda))$ be a given matrix. Find a $d\times d$ matrix $\hat S$ minimizing $||\text{vec}(\hat S)\text{vec}(\hat S)^\dagger-\hat D||$.
\end{problem}
\renewcommand{\theproblem}{\arabic{problem}}
\addtocounter{problem}{-1}
\renewcommand{\theproblem}{\arabic{problem}$.2$}
\begin{problem}\label{subproblem2}
Let $\hat S$ be given. Find a $d\times d$ unitary matrix $\hat G$ minimizing $||\text{vec}(\hat G)\text{vec}(\hat G)^\dagger-\text{vec}(\hat S)\text{vec}(\hat S)^\dagger||$.
\end{problem}
\renewcommand{\theproblem}{\arabic{problem}}

\subsubsection{}
For Problem \ref{subproblem1}, let
\begin{equation*}
\begin{aligned}
L_1& =||\text{vec}(\hat S)\text{vec}(\hat S)^{\dagger}-\hat D||^2\\
& =\text{Tr}\{[\text{vec}(\hat S)\text{vec}(\hat S)^\dagger-\hat D][\text{vec}(\hat S)\text{vec}(\hat S) ^\dagger-\hat D^\dagger]\}\\
& =[\text{vec}(\hat S)^\dagger\text{vec}(\hat S)]^2-\text{vec}(\hat S)^\dagger(\hat D+\hat D^\dagger)\text{vec}(\hat S) +\text{Tr}(\hat D\hat D^\dagger).\\
\end{aligned}
\end{equation*}
Then by partial differentiation we obtain the conjugate gradient matrix
\begin{equation}
\frac{\partial L_1}{\partial \text{vec}(\hat S)^*}=2\text{vec}(\hat S)^{\dagger}\text{vec}(\hat S)\text{vec}(\hat S)-(\hat D^{\dagger}+\hat D)\text{vec}(\hat S),\\
\end{equation}
which leads to
\begin{equation}
(\hat D^{\dagger}+\hat D)\text{vec}(\hat S)=2\text{vec}(\hat S)^\dagger \text{vec}(\hat S)\text{vec}(\hat S).
\end{equation}
Therefore the optimal $\text{vec}(\hat S)$ must be an eigenvector of $(\hat D^{\dagger}+\hat D)$ corresponding to the positive eigenvalue $2\text{vec}(\hat S)^\dagger \text{vec}(\hat S)$. Then
\begin{equation*}
\begin{aligned}
L_1& =[\text{vec}(\hat S)^\dagger \text{vec}(\hat S)]^2-\text{vec}(\hat S)^\dagger(\hat D+\hat D^\dagger)\text{vec}(\hat S)+\text{Tr}(\hat D\hat D^\dagger)\\
& =\text{Tr}(\hat D\hat D^\dagger)-[2\text{vec}(\hat S)^\dagger \text{vec}(\hat S)]^2/4.\\
\end{aligned}
\end{equation*}

Since $\hat D^\dagger+\hat D$ is Hermitian, we have the spectral decomposition
\begin{equation}\label{spec1}
\hat D^\dagger+\hat D=\sum_{i=1}^{d^2}\hat \alpha_i\text{vec}(\hat P_i)\text{vec}(\hat P_i)^\dagger ,
\end{equation}
where $\hat P_i\in\mathbb{C}_{d\times d}$ and $\hat \alpha_1\geq...\geq\hat \alpha_{d^2}$. To minimize $L_1$, we should choose $2\text{vec}(\hat S)^\dagger \text{vec}(\hat S)=\hat\alpha_1$ and $\hat S=\sqrt{\frac{\hat \alpha_1}{2}}\hat P_1$.

\subsubsection{}
For Problem \ref{subproblem2}, note that
\begin{equation*}
\begin{aligned}
& \ \ \ \ ||\text{vec}(\hat G)\text{vec}(\hat G)^{\dagger}-\text{vec}(\hat S)\text{vec}(\hat S)^{\dagger}||^2\\
& =\text{Tr}\{[\text{vec}(\hat G)\text{vec}(\hat G)^{\dagger}-\text{vec}(\hat S)\text{vec}(\hat S)^{\dagger}]^2\}\\
& =[\text{vec}(\hat G)^\dagger\text{vec}(\hat G)]^2+[\text{vec}(\hat S)^\dagger\text{vec}(\hat S)]^2\\
&\ \ \ \ \ \ -2\text{vec}(\hat G)^\dagger\text{vec}(\hat S)\text{vec}(\hat S)^\dagger \text{vec}(\hat G)\\
& =d^2+[\text{Tr}(\hat S^\dagger\hat S)]^2-2|\text{Tr}(\hat G^\dagger\hat S)|^2.\\
\end{aligned}
\end{equation*}
Hence, Problem \ref{subproblem2} is equivalent to maximizing $L_2=|\text{Tr}(\hat G^\dagger\hat S)|^2$ among all unitary $\hat G$. We make a polar decomposition \cite{bhatia} of $\hat S$ to obtain $\hat S=\hat V\hat Q$, where $\hat V=\hat S(\hat S^\dagger\hat S)^{-\frac{1}{2}}$ is unitary and $\hat Q=(\hat S^\dagger\hat S)^{\frac{1}{2}}$ is positive semidefinite. We make a spectral decomposition on $\hat Q$ to obtain $\hat Q=\hat Z \hat R\hat Z^\dagger$, where $\hat Z$ is unitary and $\hat R=\text{diag}(\hat R_{11},\hat R_{22},...,\hat R_{dd})$ with $\hat R_{jj}\geq0$. Without loss of generality, we assume $\hat R_{jj}>0$ for all $1\leq j\leq d$. Let $\hat F=\hat Z^\dagger\hat G^\dagger\hat V\hat Z$, and assume that $\hat F_{jj}=\hat r_je^{i\hat\psi_j}$ with $\hat r_j\geq0$ and $0\leq \hat\psi_j<2\pi$. Because $\hat F$ is unitary, we must have $\hat r_j\leq1$. Hence,
we have
\begin{equation}
\begin{array}{rl}
L_2&=|\text{Tr}(\hat G^\dagger\hat V\hat Q)|^2\\
&=|\text{Tr}(\hat F\hat R)|^2\\
&=|\sum_{j}\hat R_{jj}\hat r_je^{i\hat\psi_j}|^2\\
&=(\sum_{j}\hat r_j\hat R_{jj}\cos\hat\psi_j)^2+(\sum_{j}\hat r_j\hat R_{jj}\sin\hat\psi_j)^2.\\
\end{array}
\end{equation}
Then we let $\frac{\partial L_2}{\partial \hat\psi_j}=0$ for all $j$ and we obtain $$\frac{\sum_{j}\hat r_j\hat R_{jj}\sin\hat\psi_j}{\sum_{j}\hat r_j\hat R_{jj}\cos\hat\psi_j}=\tan\hat\psi_1=\tan\hat\psi_2=...=\tan\hat\psi_d.$$ Note that $L_2(\hat F)=L_2(e^{i\hat\psi_0}\hat F)$ for any $\hat\psi_0\in\mathbb{R}$. Hence, we can choose $\hat\psi_1=0$, which means $\hat\psi_j=0$ or $\pi$ for $2\leq j\leq d$. To maximize $L_2$, we should let all $\hat\psi_j$ equal to $0$. Therefore, $L_2=(\sum_{j}\hat r_j\hat R_{jj})^2$, which indicates $\hat r_j=1$ for all $j$. If all the diagonal elements of a unitary matrix are equal to one, then it must be the identity matrix. Hence, for the optimal value we have $\hat F=I$. Considering an extra global phase, we finally have the optimal solution $$\hat G=e^{i\hat\psi}\hat V=e^{i\hat\psi}\hat S(\hat S^\dagger \hat S)^{-\frac{1}{2}},$$ where $\hat\psi\in\mathbb{R}$. Combining the results of Problem \ref{subproblem1} and Problem \ref{subproblem2}, we obtain the final solution.

After we solve Problem \ref{problem3}, we should calculate the Kraus operator $\hat A$ (which is also the unitary propagator $\hat U(t)$) from $\hat G$, and finally we calculate $\hat H$ from $\hat A$. Note that $\hat U(t)$ must be a unitary matrix. Then the questions arise of how to calculate $\hat A$ from $\hat G$, and whether the matrix $\hat A$ calculated from $\hat G$ is always unitary? We answer these questions as follows.

\begin{proposition}\label{unitary1}
Under the assumptions of Problem \ref{problem3}, suppose we have obtained a solution $$\hat X=\text{vec}(\hat G)\text{vec}(\hat G)^{\dagger}.$$ Then there is essentially only one Kraus operator $\hat A$ calculated from $\hat G$. $\hat A$ must be unitary and in fact $\hat A$ is equal to $e^{i\phi} \hat G^T$, where $\phi \in \mathbb{R}$.
\end{proposition}
\begin{IEEEproof}
Denote $\text{vec}(\hat G)_j$ as the $j$-th element of $\text{vec}(\hat G)$. Since $$\hat {X}=\text{vec}(\hat G)\text{vec}(\hat G)^{\dagger},$$ then
\begin{equation}
\begin{array}{rl}
\hat{\mathcal{E}}(\rho)
=&\!\!\!\!\sum_{j,k=1}^{d^2}E_j\rho E_k^\dagger \hat x_{jk}\\
=&\!\!\!\!\sum_{j,k=1}^{d^2}E_j\rho E_k^\dagger[\text{vec}(\hat G)_j\text{vec}(\hat G)_k^*]\\
=&\!\!\!\!\sum_{j=1}^{d^2}E_j\text{vec}(\hat G)_j\rho\sum_{k=1}^{d^2}E_k^\dagger \text{vec}(\hat G)_k^*\\
=&\!\!\!\!\sum_{m,n=1}^d|m\rangle\langle n|\text{vec}(\hat G)_{(m-1)d+n}\rho\\
& \times\sum_{s,t=1}^d|t\rangle\langle s|\text{vec}(\hat G)_{(s-1)d+t}^*\\
=&\!\!\!\!\sum_{m,n=1}^d|m\rangle\langle n|\hat G_{nm}\rho\sum_{s,t=1}^d|t\rangle\langle s|\hat G_{ts}^*\\
=&\!\!\!\!\hat G^T\rho \hat G^*\\
=&\!\!\!\!e^{i\phi}\hat G^T\rho e^{-i\phi}\hat G^*.\\
\end{array}
\end{equation}
Therefore, there is essentially only one Kraus operator, which is $e^{i\phi}\hat G^T$ with $\phi\in\mathbb{R}$ undetermined, and $\hat A=e^{i\phi}\hat G^T$ is unitary.
\end{IEEEproof}

\begin{remark}
If $\hat S$ and $\hat G$ are the solutions to Problem \ref{subproblem1} and Problem \ref{subproblem2}, respectively, then for any $\phi_1,\phi_2\in\mathbb{R}$, $e^{i\phi_1}\hat S$ and $e^{i\phi_2}\hat G$ are also optimal solutions, respectively. Hence, there is in fact an undetermined global phase in $\hat G$, which can also be seen from Proposition \ref{unitary1}. This stems from the global phase in the Hamiltonian, which is physically unobservable. Through proper prior knowledge, this global phase can be eliminated. For example, in \cite{zhang 2014} the prior knowledge of $\text{Tr}H=0$ is assumed. In our simulations of Section \ref{secsim}, we use the assumption that the smallest eigenvalue of $H$ is set to a determined value.
\end{remark}

After obtaining $\hat A$, we need to solve $\hat A=e^{-i\hat Ht}$ to obtain $\hat H$. Note that in real physical systems we always require $\hat H$ to be Hermitian. Another question which naturally arises is whether every solution $\hat H$ of the equation $\hat A=e^{-i\hat Ht}$ is Hermitian? We introduce Theorem 1.43 from \cite{higham} as well as its proof, since the proof provides a method to obtain $\hat H$.
\begin{lemma}[\cite{higham}]\label{schur1}
$A\in \mathbb C_{n\times n}$ is unitary if and only if $A=e^{iH}$ for some Hermitian $H$. In this representation $H$ can be taken to be Hermitian positive definite.
\end{lemma}
\begin{IEEEproof}
The Schur decomposition of $A$ has the form $A=QDQ^\dagger$ with $Q$ unitary and $$D=\text{diag}(\text{exp}(i\theta_j))=\text{exp}(i\Theta),$$ where $\Theta=\text{diag}(\theta_j)\in\mathbb R_{n\times n}$. Hence, $$A=Q\text{exp}(i\Theta)Q^\dagger=\text{exp}(iQ\Theta Q^\dagger)=\text{exp}(iH),$$ where $H=H^\dagger$. Without loss of generality we can take $\theta_j>0$, which implies that $H$ is positive definite.
\end{IEEEproof}

Lemma \ref{schur1} satisfies our needs perfectly. Instead of using the general matrix logarithm function, we can just use the Schur decomposition to obtain the logarithm of unitary matrix $\hat A$. Furthermore, from the proof of Lemma \ref{schur1} we notice that all $\theta_j$ should lie in a region no larger than $\pi$, otherwise they can not be uniquely determined. This indicates that the sampling period should be small enough. This can also be viewed as a result of Nyquist sampling theorem, as stated in \cite{zhang 2014}. Hence, in this paper we employ the following assumption.
\begin{assumption}\label{assump}
The evolution time $t$ satisfies
\begin{equation}
0<t<\frac{\pi}{h_d-h_1},
\end{equation}
where $h_d$ and $h_1$ are the largest and smallest eigenvalues of Hamiltonian $H$, respectively.
\end{assumption}

In Appendix \ref{appendix1} we give an example of a sufficient condition for Assumption \ref{assump}, which might be more convenient to determine $t$ in practice. Now with Assumption \ref{assump} satisfied and $h_1$ set, we design an algorithm to recover the Hamiltonian from a unitary $\hat G$ as the following.

\begin{algorithm}\label{alg3}
(i) Perform a Schur decomposition of $\hat G^T$ to get $\hat G^T=\hat Q\hat J\hat Q^\dagger$ with $\hat Q$ unitary, and $\hat J=\text{exp}(i\hat\Theta)$, where $\hat\Theta=\text{diag}(\hat\theta_j)$, $0\leq \hat\theta_1\leq\hat\theta_2\leq ...\leq\hat\theta_d<2\pi$.

(ii) If $\hat\theta_d-\hat\theta_1< \pi$, go to step (iii); otherwise, find the smallest $k$ so that $\hat\theta_k-\hat\theta_1\geq\pi$. Then for $j=k,k+1,...,d$, replace $\hat\theta_j$ with $\hat\theta_j-2\pi$. This step aims to ensure the reconstructed Hamiltonian has spectral region no larger than $h_d-h_1$.

(iii) Let $\hat\theta_0=\max_j{\hat\theta_j}$. For all $1\leq j\leq d$, take $\bar{\theta}_j=\hat\theta_j-h_1t-\hat\theta_0$. If we denote $\bar{\Theta}=\text{diag}(\bar{\theta}_j)$, then $\hat H=-Q\bar{\Theta} Q^\dagger/t$ is the final estimated Hamiltonian.
\end{algorithm}

\subsection{General Procedure and Computational Complexity}

\begin{figure}
\centering
\includegraphics[width=3in]{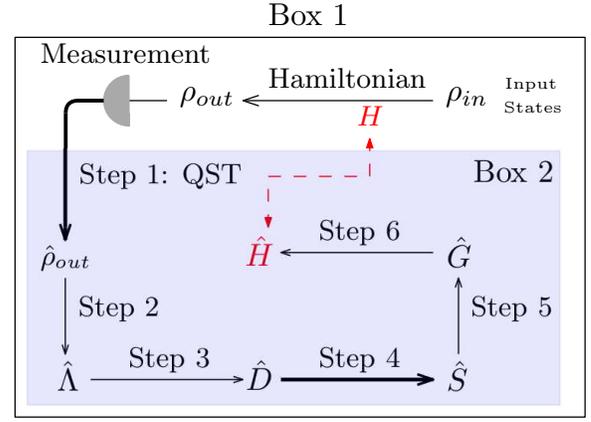}
\centering{\caption{General procedure of the quantum Hamiltonian identification method, where QST indicates quantum state tomography and this paper focuses on Box 2.}\label{procedure}}
\end{figure}

In Fig. \ref{procedure}, we summarize the general procedure of the QHI framework. All steps in Box 2 are data processing steps performed on a computer. Step 1 is quantum state tomography, which includes the acquisition of experimental data and post-processing of the experimental data. In this paper we do not consider the time spent on experiments, since it depends on the experimental realization. In the following, we briefly summarize each step and illustrate their corresponding computational complexity.

\textbf{Step 1}. Choose basis sets $\{E_i\}$ and $\{\rho_m\}$ and calculate $B$. Then use quantum state tomography to reconstruct experimental output states of the system. The number of resource copies $N$ in state tomography determines the estimation error, but does not affect the computational complexity of the estimation algorithm. Generally the calculation of $B$ according to (\ref{betadef}) has $O(d^{11})$ computational complexity. However, under the natural basis, this complexity can be reduced to only $O(d^4)$. For state reconstruction, we employ the method of quantum state tomography using LRE for our numerical simulations. The computational complexity of LRE state tomography is $O(d^6)$ offline and $O(d^4)$ online \cite{qi 2013}. Considering there are $d^2$ output states to be reconstructed, the total computational complexity of our LRE method   for QHI is $O(d^6)$.

\textbf{Step 2}. Use (\ref{eqlambda}) to determine $\hat\Lambda$. Generally the computational complexity to solve (\ref{eqlambda}) is $O(d^{12})$. But it is only $O(d^2)$ using the orthogonal property under the natural basis.

\textbf{Step 3}. Calculate $\hat D=\text{vec}^{-1}(B^\dagger\text{vec}(\hat\Lambda))$. Generally the complexity is $O(d^8)$. But under the natural basis, we already know the specific structure and value of $B$ (see (\ref{b1})). Thus, the complexity now is only $O(d^4)$.

\textbf{Step 4}. Calculate $\hat S$ according to the spectral decomposition of $\hat D+\hat D^\dagger$. The computational complexity is determined by spectral decomposition, which is $O(d^6)$ (the computational complexity of spectral decomposition is cubic in a Hermitian matrix's dimension, see \cite{schur2}).

\textbf{Step 5}. Use matrix polar decomposition to obtain $\hat G=\hat S(\hat S^\dagger\hat S)^{-\frac{1}{2}}$. The computational complexity is $O(d^3)$ \cite{schur2}.

\textbf{Step 6}. Use the Schur decomposition to obtain the final estimated Hamiltonian $\hat H$ from $\hat G$. The computational complexity of Schur decomposition is $O(d^3)$ \cite{schur2,schur1}.

Our Hamiltonian identification procedure has the following advantages. Firstly, the framework is general, since we formulate it within the QPT framework. We do not impose any restriction (such as sparseness) on the Hamiltonian. Secondly, Step 1 has the potential for parallel processing. One can deal with data on hand to reconstruct existing output states while at the same time inputting new probe states to the process and making measurements on them. Thirdly, the computational complexity can be analyzed. Regardless of the time spent in experiments, all steps in our QHI framework have clear computational complexity (at most $O(d^6)$). Finally, it is possible to analytically investigate an error upper bound and a detailed error analysis is presented in Section \ref{err}.

\subsection{Practical Consideration of Storage Requirements}\label{bcal}
One issue in the calculations is that the dimension of $B$ may increase rapidly. When there are 4 qubits, $B$ has $2^{32}$ elements. If it takes one byte to store one element of $B$, then we need $4GB$ of storage space, which is already a very heavy task for a common PC. We notice that $B$ generated from the natural basis is a permutation matrix. This is vital to computation efficiency. A permutation matrix is a (square in this paper) matrix such that all elements are $0$ except exactly one $1$ in each column and each row.

Notice that after $B$ is determined from equations (\ref{betadef}) and (\ref{matrixB}), its real usage is in Problem \ref{problem3}, where we need to multiply $B^\dagger$ to a vector. This multiplication task can be done in an alternative way where $B$'s full storage is avoided. To be specific, we aim to make $B$ sparse. Hence, we only need to store the information of its very small number of nonzero elements and thus ignore a large number of zero elements, while still being able to perform the multiplication. This idea is realized by the following theorem:

\begin{theorem}\label{Bsparse}
Let $\{E_i\}_{i=1}^{d^2}$ be a set of matrices in $\mathbb C_{d\times d}$. Choose $\{\rho_m\}_{m=1}^{d^2}=\{|j\rangle\langle k|\}_{1\leq j,k\leq d}$. Define $B$ through (\ref{betadef}) and (\ref{matrixB}). Then $\{E_i\}=e^{i\theta}\{\rho_m\}$ if and only if $B$ is a permutation matrix. Here, $\theta \in \mathbb{R}$ is any fixed global phase.
\end{theorem}

\begin{IEEEproof}
Using Theorem \ref{proequiv} we know that equation (\ref{wexpress1}) holds.

$Sufficiency:$ Define $W(j,k)$ as a $d^{2}\times d^{2}$ matrix where $W(j,k)$'s element in position $(m,n)$ is the number $\beta_{nm}^{jk}$, and denote $(x,y)=(x-1)K+y$ for $1\leq x,y\leq K$. Using equation (\ref{property6}), we consider each element of $W(j,k)$,
\begin{equation}\label{eqq3}
\begin{array}{rl}
&W(j,k)_{(p,q)(s,t)}\\
=&\text{vec}(\rho_{(p,q)})^\dagger(e^{-i\theta}|g\rangle^*\langle h|^*\otimes e^{i\theta}|m\rangle\langle n|)\text{vec}(\rho_{(s,t)})\\
=&(\langle q|^*\otimes\langle p|)(|g\rangle^*\otimes|m\rangle)(\langle h|^*\otimes\langle n|)(|t\rangle^*\otimes|s\rangle)\\
=&(\langle q|g\rangle^*\otimes\langle p|m\rangle)(\langle h|t\rangle^*\otimes\langle n|s\rangle)\\
=&\delta_{qg}\delta_{pm}\delta_{th}\delta_{sn}.\\
\end{array}
\end{equation}
Hence, each matrix $W(j,k)$ has exactly one 1 and all other elements are 0. From equation (\ref{bexpress1}) we know each row of $B$ has exactly one 1 and all other elements are 0. When indices $j$ and $k$ run from $1$ to $d^2$, the index combination $(g,h,m,n)$ never repeats, therefore, $W(j_1,k_1)$ and $W(j_2,k_2)$ have different positions of 1 as long as index pair $(j_1,k_1)\neq(j_2,k_2)$. This means each row of $B$ has no more than one 1. Since $B$ is square, we know that each row of $B$ has exactly one 1. Hence, $B$ is a permutation matrix.

$Necessity:$ When $B$ is a permutation matrix, from equation (\ref{bexpress1}) we know that each matrix $W(j,k)$ has exactly one 1 and all other elements are 0. According to $V$'s permutation property and equation (\ref{wexpress2}) we know this property for each $W(j,k)$ also holds for each matrix $E_k^*\otimes E_j$. This means that each matrix $E_j$ has exactly one nonzero element, denoted as $x_j$. Then we have $x_k^*x_j=1$ holds for every $k,j=1,2,...,d^2$. Let $j=k$, and we find $x_j=e^{i\theta_j}$. Then we know $\theta_1=\theta_2=...=\theta_{d^2}=\theta$, where $\theta$ is any fixed real number. Since $B$ is invertible, from Theorem \ref{binvert} we know $\{E_j\}$ is a linearly independent set. Thus each pair of matrices in $\{E_j\}$ have different positions of $e^{i\theta}$. Hence, we can write $\{E_j\}=e^{i\theta}\{\rho_m\}$.
\end{IEEEproof}

From the proof of Theorem \ref{Bsparse}, one can also deduce an equation to directly calculate $B$. Substituting this into equation (\ref{eqq3}), we obtain
\begin{equation}\label{b1}
\beta^{jk}_{(s,t)(p,q)}=\delta_{qg}\delta_{pm}\delta_{th}\delta_{sn}=\beta^{(m,n)(g,h)}_{(s,t)(p,q)}.
\end{equation}
Therefore, one can easily write down $B$ when the size $d$ is given.

A special case of the sufficiency of Theorem \ref{Bsparse}; i.e., when $\{E_i\}$ and $\{\rho_m\}$ are the same natural basis sets with the same order of elements, also appeared in \cite{wuxiaohua}. Our theorem and proof here is more general. Using this theorem, we only need to store all $1$'s positions in $B$, which only requires $d^4$ storage space. This is a great reduction compared with $d^8$, and the cost is only some more coding in calculating multiplication by $B$. Furthermore, the computational complexity in writing down $B$ is also reduced to only $O(d^4)$.

\section{ERROR ANALYSIS}\label{err}
The error in the Hamiltonian identification method under consideration has only three possible sources. The first one occurs in state estimation, where measurement frequency in practical simulations or experiments is used to approximate the measurement probability. The second one is that state reconstruction algorithm might produce errors. The third one is that our TSO QHI algorithm may also produce errors. In this section, we give an error upper bound. We first fix the given evolution time $t$ and analyze the error of our QHI method. Then we utilize the similar method to analyze the relationship between the error and the time $t$.
\subsection{Upper Error Bound for Fixed Evolution Time}
\begin{theorem}\label{mainthe}
If $\{E_i\}$ and $\{\rho_m\}$ are chosen as natural basis of $\mathbb{C}_{d\times d}$ and the evolution time $t$ is fixed and satisfies Assumption \ref{assump}, then the estimation error of the TSO QHI method $E||\hat H-H||$ scales as $O(\frac{d^3}{\sqrt{N}})$, where $N$ is the number of resources in state tomography for each output state.
\end{theorem}

\begin{IEEEproof}
The proof of this theorem is divided into the following seven parts.
\subsubsection{Error in step 1}
The quantum state tomography algorithm used in this paper is from \cite{qi 2013}, and the upper bound on the state estimation error is given by
\begin{equation}\label{stateupper}
\sup_\rho E\text{Tr}(\hat\rho-\rho)^2=\frac{M}{4N}\text{Tr}(X^TX)^{-1},
\end{equation}
where $\rho$ is the true state and $\hat\rho$ its estimator, $M$ is the number of measurement bases, $N$ is the number of experiments (i.e., number of copies of $\rho$) in state tomography, $X$ is a matrix determined by the measurement basis set (for details, see \cite{qi 2013}). Henceforth, we denote this error upper bound (i.e., the RHS of (\ref{stateupper})) as $\Delta_{st}$. Following the deduction in the Methods section of \cite{qi 2013}, one can prove $\Delta_{st}\sim O(\frac{d^4}{N})$. In the following, we will label other errors in the form of $\Delta$ with a subscript.

When $\rho_m$ is Hermitian, $E||\hat{\mathcal{E}}(\rho_m)-\mathcal{E}(\rho_m)||^2\leq \Delta_{st}$. When $\rho_m$ is not Hermitian, its process output is in fact calculated according to equation (\ref{eq10}) rather than directly probed. Hence, we must analyze this situation specifically. Under the choice of $\{\rho_m\}$ as the natural basis, for $j\neq k$,
\begin{equation}
\begin{array}{rl}
&E||\hat{\mathcal{E}}(|j\rangle\langle k|)-\mathcal{E}(|j\rangle\langle k|)||^2 \\
=& E||[\hat{\mathcal{E}}(|+\rangle\langle +|)-\mathcal{E}(|+\rangle\langle +|)]+i[\hat{\mathcal{E}}(|-\rangle\langle -|)-\mathcal{E}(|-\rangle\langle -|)]\\
&\ \ -\frac{1+i}{2}[\hat{\mathcal{E}}(|j\rangle\langle j|)-\mathcal{E}(|j\rangle\langle j|)]-\frac{1+i}{2}[\hat{\mathcal{E}}(|k\rangle\langle k|)-\mathcal{E}(|k\rangle\langle k|)]||^2\\
\leq & (1+|i|+|\frac{1+i}{2}|+|\frac{1+i}{2}|)^2\Delta_{st}\\
= & (6+4\sqrt{2})\Delta_{st}.\\
\end{array}
\end{equation}

\subsubsection{Error in step 2}
Now we calculate the error in the experimental data:
\begin{equation}\label{lambupper}
\begin{array}{rl}
&E||\hat\Lambda-\Lambda||^2 \\
=&E\sum_m \sum_{n,k}(\hat\lambda_{mn}^*-\lambda_{mn}^*)(\hat\lambda_{mk}-\lambda_{mk})\delta_{nk}\\
=&E\sum_m\sum_{n,k}(\hat\lambda_{mn}^*-\lambda_{mn}^*)(\hat\lambda_{mk}-\lambda_{mk})\text{Tr}(\rho_n^\dagger\rho_k)\\
=&E\sum_m\text{Tr}[\sum_n(\hat\lambda_{mn}^*-\lambda_{mn}^*)\rho_n^\dagger\sum_k(\hat\lambda_{mk}-\lambda_{mk})\rho_k]\\
=&E\sum_m\text{Tr}(\widehat{\mathcal{E}(\rho_m)}-\mathcal{E}(\rho_m))^2 \\
=&E[\sum_{j=1}^d\sum_{k=1,k\neq j}^d||\hat{\mathcal{E}}(|j\rangle\langle k|)-\mathcal{E}(|j\rangle\langle k|)||^2\\
&+\sum_{l=1}^d||\hat{\mathcal{E}}(|l\rangle\langle l|)-\mathcal{E}(|l\rangle\langle l|)||^2]\\
\leq &(6+4\sqrt{2})d(d-1)\Delta_{st}+d\Delta_{st}.\\
\end{array}
\end{equation}
Also, we denote $\Delta_\Lambda=||\hat\Lambda-\Lambda||$.

\subsubsection{Error in step 3}
From Theorem \ref{Bsparse}, we know $B^\dagger$ is a permutation matrix. Hence, its effect on $\text{vec}(\Lambda)$ is merely a series of interchanging two elements of $\text{vec}(\Lambda)$, and thus $D=\text{vec}^{-1}(B^\dagger \text{vec}(\Lambda))$ is just a reordering of $\Lambda$'s elements. For the same reason, $\hat D-D$ is just reordering of $\hat\Lambda -\Lambda$. Therefore
\begin{equation}\label{eqtemx1}
||\hat D-D||=(\sum_{j,k}|\hat D_{jk}-D_{jk}|^2)^{\frac{1}{2}}=||\hat \Lambda-\Lambda||=\Delta_\Lambda.
\end{equation}

\subsubsection{Error in step 4}
We present a lemma to be used in this part.
\begin{lemma}\label{lemma2}
Let $b$ and $c$ be two complex vectors with the same finite dimension and assume that they are not both zero simultaneously. Then we have
\begin{equation}
\frac{||bb^\dagger-cc^\dagger||}{||b||+||c||}\leq \min_{\theta\in\mathbb R} ||e^{i\theta}b-c||\leq\frac{\sqrt{2}||bb^\dagger-cc^\dagger||}{\sqrt{||b||^2+||c||^2}}.
\end{equation}
\end{lemma}
The detailed proof of Lemma \ref{lemma2} can be found in Appendix \ref{app4}.

We first estimate $||\hat S||$. $$||\hat S||^2=\text{Tr}(\hat S^\dagger \hat S)=\frac{\hat\alpha_1}{2}\text{Tr}(\hat P_1^\dagger \hat P_1)=\frac{\hat\alpha_1}{2}.$$ Remember that $\hat\alpha_1$ is the largest eigenvalue of $\hat D+\hat D^\dagger$. Using Theorem \ref{weyl}, $$|\hat\alpha_1-2d|\leq||(\hat D+\hat D^\dagger)-2D||\leq2||\hat D-D||=2\Delta_\Lambda.$$ We thus have $$2d-2\Delta_\Lambda\leq \hat\alpha_1\leq2d+2\Delta_\Lambda.$$ Therefore,
\begin{equation}\label{eqtem4}
\sqrt{d-\Delta_\Lambda}\leq||\hat S||=\sqrt{\hat \alpha_1/2}\leq\sqrt{d+\Delta_\Lambda}.
\end{equation}

We also need to estimate $||\hat S-S||$. Using Lemma \ref{lemma2}, we have
\begin{equation}\label{eqtem1}
\begin{array}{rl}
&||\hat S-S||=||\text{vec}(\hat S)-\text{vec}(S)||\\
\leq&\frac{\sqrt{2}||\text{vec}(\hat S)\text{vec}(\hat S)^\dagger-\text{vec}(S)\text{vec}(S)^\dagger||}{\sqrt{||\text{vec}(\hat S)||^2+||\text{vec}(S)||^2}}\\
\leq&\frac{\sqrt{2}||\text{vec}(\hat S)\text{vec}(\hat S)^\dagger-\text{vec}(S)\text{vec}(S)^\dagger||}{\sqrt{2d-\Delta_\Lambda}}\\
=&[\frac{1}{\sqrt{d}}+o(1)]||\text{vec}(\hat S)\text{vec}(\hat S)^\dagger-D||\\
\leq&[\frac{1}{\sqrt{d}}+o(1)][||\text{vec}(\hat S)\text{vec}(\hat S)^\dagger-\hat D||+||\hat D-D||]\\
=&[\frac{1}{\sqrt{d}}+o(1)][\min_{\tilde{S}\in\mathbb{C}_{d\times d}}||\text{vec}(\tilde S)\text{vec}(\tilde S)^\dagger-\hat D||+||\hat D-D||]\\
\leq&[\frac{1}{\sqrt{d}}+o(1)][||\text{vec}(G)\text{vec}(G)^\dagger-\hat D||+||\hat D-D||]\\
=&[\frac{1}{\sqrt{d}}+o(1)]\cdot 2||\hat D-D||\sim \frac{2}{\sqrt{d}}\Delta_\Lambda.\\
\end{array}
\end{equation}

\subsubsection{Error in step 5}
We introduce \textit{Weyl's Perturbation Theorem}, which can be found in \cite{bhatia}.
\begin{lemma}[\cite{bhatia}]\label{weyl}
Let $A$, $B$ be Hermitian matrices with eigenvalues $\lambda_1(A)\geq...\geq\lambda_n(A)$ and $\lambda_1(B)\geq...\geq\lambda_n(B)$, respectively. Then
\begin{equation}
\max_j|\lambda_j(A)-\lambda_j(B)|\leq||A-B||.
\end{equation}
\end{lemma}

\begin{remark}
The original version of Lemma \ref{weyl} was for the operator norm. However, from \cite{bhatia} we know for any finite-dimension square matrix, its operator norm is not larger than its Frobenius norm. Therefore this theorem also holds for the Frobenius norm, which is our main focus throughout this paper.
\end{remark}

For the true value we have $S^\dagger S=G^\dagger G=I$ and $||S||=\sqrt{d}$. Denote $||\hat S^\dagger \hat S-S^\dagger S||=\Delta_{S^\dagger S}$. From the spectral decomposition $\hat S^\dagger\hat S=\hat U\hat E\hat U^\dagger$, where $\hat E=\text{diag}(1+t_1,1+t_2,...1+t_d)$. Hence, $t_j\in\mathbb R$. Then $$||\hat S^\dagger \hat S-S^\dagger S||^2=||\hat U\hat E\hat U^\dagger-I||^2=||\hat E-I||^2=\sum_j t_j^2=\Delta_{S^\dagger S}^2.$$ Thus we know
\begin{equation}\label{eqtem6}
\begin{array}{rl}
||\hat{G}-\hat{S}||^2=&\text{Tr}[(\hat{G}^\dagger-\hat{S}^\dagger)(\hat{G}-\hat{S})]\\
=&d-2\text{Tr}\sqrt{\hat S^\dagger \hat S}+\text{Tr}(\hat S^\dagger \hat S)\\
=&d-2\sum_j\sqrt{1+t_j}+\sum_j(1+t_j)\\
=&\sum_j(\sqrt{1+t_j}-1)^2=\sum_j\frac{t_j^2}{2+t_j+2\sqrt{1+t_j}}\\
=&\sum_j t_j^2[\frac{1}{4}-\frac{1}{8}t_j+o(t_j)]=\frac{1}{4}\Delta_{S^\dagger S}^2+o(\Delta_{S^\dagger S}^2).\\
\end{array}
\end{equation}

For $\Delta_{S^\dagger S}$, using property (\ref{propertyhs2}), we have
\begin{equation}\label{eqtem3}
\begin{array}{rl}
\Delta_{S^\dagger S}=&||\hat S^\dagger \hat S-S^\dagger S||\\
\leq&||\hat S^\dagger \hat S-\hat S^\dagger S||+||\hat S^\dagger S- S^\dagger S||\\
\leq&||\hat S^\dagger||\cdot||\hat S-S||+||S||\cdot||\hat S^\dagger- S^\dagger||\\
=&(||\hat S||+\sqrt{d})||\hat S-S||\\
\leq&(\sqrt{d+\Delta_\Lambda}+\sqrt{d})||\hat S-S||\\
\sim&(\sqrt{d+\Delta_\Lambda}+\sqrt{d})\frac{2}{\sqrt{d}}\Delta_\Lambda\sim4\Delta_\Lambda.\\
\end{array}
\end{equation}
Combining (\ref{eqtem6}) and (\ref{eqtem3}), we obtain
\begin{equation}\label{eqtem09}
||\hat{G}-\hat{S}||\sim\frac{1}{2}\Delta_{S^\dagger S}\leq 2\Delta_\Lambda.
\end{equation}

From subsection \ref{subsec2}, we know there is in fact an extra degree of freedom $\phi$ in the estimated $e^{i\phi}\hat G^T$ and it can be eliminated using prior knowledge. Here, we take $$||\hat G-G||=\min_\phi||e^{i\phi}\hat G-G||.$$

Then we have
\begin{equation}\label{gupper}
\begin{array}{rl}
||\hat G-G||
\leq& ||\hat G-\hat S||+||\hat{S}-S||+||S-G||\\
=&||\hat G-\hat S||+||\hat{S}-S||.\\
\end{array}
\end{equation}

Now by substituting equations (\ref{eqtem1}) and (\ref{eqtem09}) into (\ref{gupper}), we have
\begin{equation}\label{gupper2}
\begin{array}{rl}
||\hat G-G||\leq & ||\hat G-\hat S||+||\hat{S}-S||\\
\leq &2\Delta_\Lambda+\frac{2}{\sqrt{d}}\Delta_\Lambda\sim O(\Delta_\Lambda).
\end{array}
\end{equation}

\subsubsection{Error in step 6}
In this part we need the following lemma:
\begin{lemma}\label{lemma1}
For $\theta\in[-\pi,\pi]$, $\frac{2}{\pi^2}\theta^2\leq1-\cos\theta$.
\end{lemma}

Based on differential analysis up to the second-order derivative, the proof of Lemma \ref{lemma1} is straightforward and hence we omit the details.

Suppose the system Hamiltonian has a spectral decomposition $tH=-Q\Theta Q^\dagger$, where $\Theta=\text{diag}(\theta_j)$. Since $t$ satisfies Assumption \ref{assump}, we have $0\leq \theta_j\leq \pi$ for every $j=1,2,...,d$. Let $M=\hat Q^\dagger Q$, which is also unitary. Then, we have
\begin{equation}
\begin{array}{rl}
&t^2||\hat{H}-H||^2 \\
=&||\hat{Q}\hat\Theta\hat{Q}^\dagger-Q\Theta Q^\dagger||^2=||\hat\Theta-M\Theta M^\dagger||^2\\
=&\text{Tr}(\hat\Theta^2+\Theta^2)-2\text{Tr}(\hat\Theta M\Theta M^\dagger)\\
=&\sum_j(\hat\theta_j^2+\theta_j^2)-2\sum_{j,k}\hat\theta_j\theta_k|M_{jk}|^2\\
=&\sum_{j,k}(\hat\theta_j^2+\theta_k^2)|M_{jk}|^2-2\sum_{j,k}\hat\theta_j\theta_k|M_{jk}|^2\\
=&\sum_{j,k}(\hat\theta_j-\theta_k)^2|M_{jk}|^2.\\
\end{array}
\end{equation}

Now using Lemma \ref{lemma1}, we have
\begin{equation}\label{eq88}
\begin{array}{rl}
&\frac{4t^2}{\pi^2}||\hat{H}-H||^2\\
\leq &2\sum_{j,k}[1-\cos(\hat\theta_j-\theta_k)]|M_{jk}|^2\\
=&2\sum_{j,k}|M_{jk}|^2-2\sum_{j,k}\cos(\theta_k-\hat\theta_j)|M_{jk}|^2\\
=&2d-2\text{Re}(\sum_{j,k}e^{i(\theta_k-\hat\theta_j)}|M_{jk}|^2)\\
=&\text{Tr}(I+I)-2\text{Re}(\sum_{j}e^{-i\hat\theta_j}\sum_k e^{i\theta_k}|M_{jk}|^2)\\
=&\text{Tr}(e^{i\hat\Theta}e^{-i\hat\Theta}+e^{i\Theta}e^{-i\Theta})-2\text{Re}[\text{Tr}(e^{-i\hat\Theta}M e^{i\Theta}M^\dagger)]\\
=&||e^{i\hat\Theta}-M e^{i\Theta}M^\dagger||^2=||\hat{Q}e^{i\hat\Theta}\hat{Q}^\dagger-Qe^{i\Theta}Q^\dagger||^2\\
=&||\hat G^T-G^T||^2=||\hat G-G||^2.\\
\end{array}
\end{equation}
Hence
\begin{equation}\label{hupper}
||\hat H-H||\sim O(||\hat G-G||)
\end{equation}

\subsubsection{Total Error}

We combine equations (\ref{hupper}), (\ref{gupper2}), (\ref{lambupper}) and (\ref{stateupper}) to obtain
\begin{equation}
\begin{array}{rl}
E||\hat H-H||^2&\sim E[O(||\hat G-G||^2)]\sim E[O(\Delta_\Lambda^2)]\\
&\sim O(d^2\Delta_{st})\sim O(\frac{d^6}{N}),\\
\end{array}
\end{equation}
which concludes the proof of Theorem \ref{mainthe}.
\end{IEEEproof}

From Theorem \ref{mainthe}, we can also obtain the following corollary.
\begin{corollary}\label{aunbiased}
If $\{E_i\}$ and $\{\rho_m\}$ are chosen as natural basis of $\mathbb{C}_{d\times d}$, and the evolution time $t$ is fixed and satisfies Assumption \ref{assump}, then the TSO Hamiltonian identification method is asymptotically unbiased.
\end{corollary}

\subsection{Upper Error Bound vs Evolution Time}
Using a similar idea to the above, we can characterize the estimation error for different evolution times $t$:
\begin{theorem}\label{mainthe2}
If $\{E_i\}$ and $\{\rho_m\}$ are chosen as a natural basis of $\mathbb{C}_{d\times d}$ and $N$ is fixed, then the estimation error of the TSO Hamiltonian identification method scales as $E||\hat H-H||\sim O(\frac{1}{t})$ where $t$ satisfies Assumption \ref{assump}.
\end{theorem}

The proof of this theorem is similar to the proof of Theorem \ref{mainthe}. Note from (\ref{eq88}), we have
\begin{equation}\label{hupper2}
\frac{2t}{\pi}||\hat H-H||\leq||\hat G-G||,
\end{equation}
which combined with (\ref{gupper2}), (\ref{lambupper}) and (\ref{stateupper}) leads to the conclusion in the theorem. It is worth pointing out that in this theorem, the evolution time cannot be arbitrarily large, rather it must be upper bounded according to Assumption \ref{assump}. Hence, this scaling only holds in a certain region.

\section{NUMERICAL RESULTS}\label{secsim}
We perform numerical simulations using MATLAB on PC. It is worth mentioning that the selection of natural bases is only a mathematical representation tool in the identification algorithm. When performing measurements on the output states, our framework is applicable to many general measurement bases, such as cube bases \cite{burgh 2008}, MUB bases \cite{wootters 1989}-\cite{miranowicz 2014}, SIC-POVMs \cite{renes 2004}, etc. In our simulations for the TSO method, we choose cube measurement bases. The single-qubit cube measurement set consists of six measurement operators: $\{\frac{I\pm\sigma_x}{2},\frac{I\pm\sigma_y}{2},\frac{I\pm\sigma_z}{2}\}$, and the multi-qubit cube measurement set is the tensor product of the single-qubit cube set. After the measurements, we then use the LRE method to reconstruct the output states.

\subsection{Performance Illustration}\label{simusec1}
\begin{figure}
\centering
\includegraphics[width=3in]{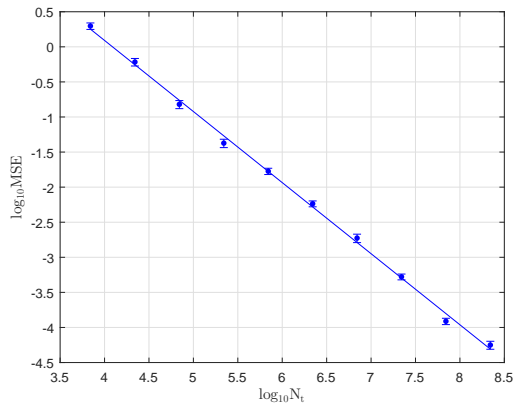}
\centering{\caption{MSE versus the logarithm of the total resource number $N_t$.}\label{test2}}
\end{figure}

First we illustrate the relationship between the Mean Squared Error (MSE) and the resource number. Let $N_t$ be the total number of resources; i.e., the total number of copies of different quantum states used as probes. Considering (\ref{eq10}), we have $N_t=\frac{3d^2-d}{2}N$. $H$ is the real Hamiltonian, and $\hat H$ its estimation through \ref{subsec2}. In Fig. \ref{test2}, the vertical axis is $\log_{10}E\text{Tr}(\hat H-H)^2$ and the horizontal axis is $\log_{10}N_t$. The real Hamiltonian is taken as
\begin{equation}\label{ham1}
H=\left(\begin{array}{*{4}{c}}
5 & 0.1 & 3i & 4i \\
0.1 & -1 & 1.8 & 0.9 \\
-3i & 1.8 & 2 & 0.7i \\
-4i & 0.9 & -0.7i & 3 \\
\end{array}\right).
\end{equation}

The distance between its largest and smallest eigenvalues is $11.95$. The evolution time $t=0.1$ and each point is repeated for 10 times. The fitting slope is $-1.0131\pm 0.0154$, which matches the theoretical result in Theorem \ref{mainthe}.

\begin{figure}
\centering
\includegraphics[width=3in]{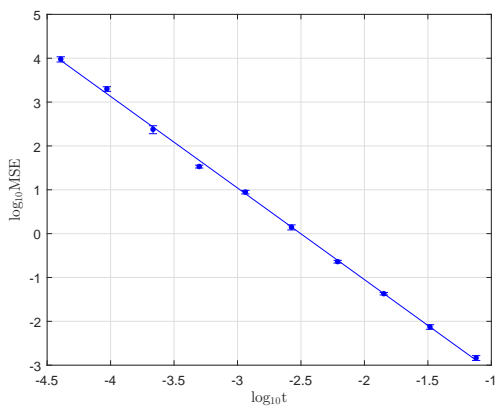}
\centering{\caption{MSE versus the logarithm of the different evolution times $t$.}\label{test5}}
\end{figure}

Now we demonstrate the relationship between the MSE and the evolution time. For the same 2-qubit Hamiltonian in (\ref{ham1}), we fix the number of copies in state tomography for each output state as $3^6\times 1000$ and perform simulations for different evolution times $t$. The result is in Fig. \ref{test5} and each point is repeated 10 times. The fitting slope is $-2.0891\pm 0.0215$, which matches the theoretical result in Theorem \ref{mainthe2}.

\begin{figure}
\centering
\includegraphics[width=3in]{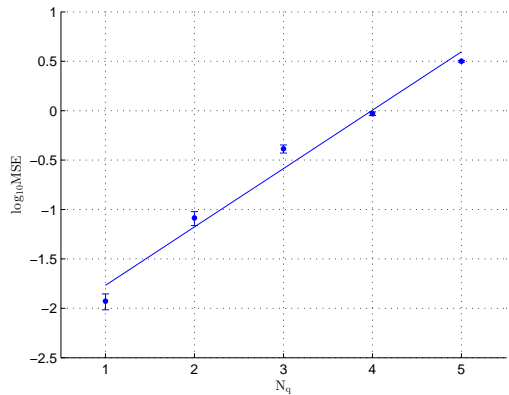}
\centering{\caption{MSE versus number of qubits $N_q$.}\label{test4}}
\end{figure}

Moreover, we present an example to illustrate the relationship between the MSE and the qubit number. Let $N_q$ denote the number of qubits; i.e., $d=2^{N_q}$. We perform simulations when $N_q$ increases from $1$ to $5$. We set $$H=\left(
\begin{array}{cc}
1 & 0.9+0.9i\\
0.9-0.9i & 2\\
\end{array}
\right)^{\otimes N_q}$$ and $t=0.01$. For $N_q=5$, the distance between the largest and smallest eigenvalues of $H$ is $193.87$. The number of copies for state tomography is $3^6\times1000$ for each output state. The result is in Fig. \ref{test4} and each point is repeated 10 times. We observe that as $N_q$ increases, the errorbar decreases. This is because as $N_q$ increases, the error is also increasing. Therefore, the fluctuations gradually become relatively small. We examine various Hamiltonians and obtain similar results. Furthermore, we observe that the upper error bound in Theorem \ref{mainthe} indicates a slope larger than that of the fitted line in Fig. \ref{test4}. The observation may come from the fact that the bound in Theorem \ref{mainthe} is an upper bound and the experimental result in Fig. \ref{test4} does not necessarily reach the bound.

\subsection{Performance Comparison}\label{simusec2}
Now we compare the performance of the TSO QHI method with the QHI approach developed by Zhang and Sarovar in \cite{zhang 2014}, which is based on the eigenstate realization algorithm in classical identification (abbreviated as the ERA method hereafter).

The ERA method can be used to give a general solution to QHI although it was originally presented for the identification of partial parameters in the system Hamiltonian. The ERA method first converts QHI into a system identification problem in the real domain, where the transfer function of the equivalent linear system can be obtained. From temporal records of system observables, it can reconstruct the transfer function. Then equating the coefficients of the transfer functions with unknown parameters to those from the experimental data, the ERA method leads to a set of multivariate polynomial equations, whose solution yields the estimates of the Hamiltonian parameters.

This approach is only efficient if the number of parameters to be identified in the Hamiltonian is small. This is because solving multivariate polynomial equation takes a considerable amount of time, especially for high dimensional systems or for full Hamiltonian identification with complex quantum systems. In fact, common algorithms solving multivariate polynomial equations can be super-exponential when the number of variables scales up~\cite{supere}.

\begin{figure}
\centering
\includegraphics[width=3in]{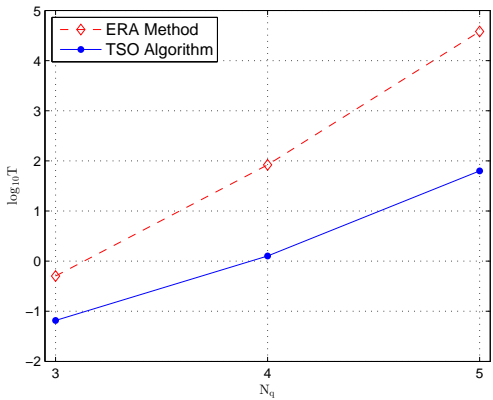}
\centering{\caption{Running time $T$ versus qubit number $N_q$ for the ERA method in \cite{zhang 2014} and our TSO method.}\label{time1}}
\end{figure}

To illustrate the efficiency of the TSO Hamiltonian identification method, we compare it with the ERA method by numerical simulations, which we performed on a single thread, computer cluster with 2 Intel Xeon E5-2680v3 CPUs and 256 GB memory. We consider the following Hamiltonian for a 1D chain of $N_q$ qubits, which is the example investigated in \cite{zhang 2014}:
\begin{equation}
H=\sum_{k=1}^{N_q}\frac{\omega_k}{2}\sigma_z^k+\sum_{k=1}^{N_q-1}\delta_k(\sigma_+^k\sigma_-^{k+1}+\sigma_-^k\sigma_+^{k+1}).
\end{equation}
Here $\omega_k$ and $\delta_k$ are unknown parameters to be identified. $\delta_k$ are the coupling strength between $k$-th and $(k+1)$-th spins, $\sigma_+=\frac{1}{2}(\sigma_x+i \sigma_y)$ and $\sigma_-=\frac{1}{2}(\sigma_x-i \sigma_y)$. Running on the same computer cluster, we compare the consumed time of our TSO QHI method versus the ERA method for the cases of $N_q=3$, $4$, and $5$. For the TSO method, we do not utilize the prior structural knowledge (1D-chain) of the targeted Hamiltonian, whereas this information is used in the ERA method. Fig. \ref{time1} shows the numerical result, where the vertical axis is the running time $T$ (in units of seconds) in a logarithmic scale, and the horizontal axis is the number of qubits $N_q$. The red diamonds are the times from the ERA method, whereas the blue dots are for the TSO identification method. The numerical results show that the TSO method is much faster (e.g., around 100 times faster for $N_q=4$) than the ERA method even if we do not use the prior knowledge of Hamiltonian's structure. It is worth mentioning that the efficiency of the TSO algorithm usually depends on the system size but not the number of parameters for a given system size, while the performance of the ERA method significantly depends on the system size as well as the number of parameters to be identified. The efficiency advantage of TSO algorithm becomes remarkable as the system size and the number of parameters to be identified increase.

\section{CONCLUSION}\label{secfinal}
We have presented a new TSO Hamiltonian identification method and analyzed its computational complexity. This identification method is applicable to general time-independent Hamiltonians for closed quantum systems. We have also provided a theoretical upper bound for the identification error and demonstrate the performance of the identification algorithm using numerical examples. Future work includes the extension of the TSO algorithm to quantum process tomography for open quantum systems and the investigation of whether quantum entanglement can enhance the performance of Hamiltonian identification.


%

\appendices
\section{PROOF OF PROPOSITION \ref{ptracepropo}}\label{app0}
\begin{IEEEproof}
In this paper, whenever we need to endow orders to number pairs $(x,y)$ ($1\leq x,y\leq K$) we identify $(x,y)$ with $(x-1)K+y$ unless declared otherwise.

When $\{E_i\}$ is chosen as $\{|j\rangle\langle k|\}_{1\leq j,k\leq d}$, expand (\ref{ptrace1}) as
\begin{equation*}
\begin{aligned}
\sum_{s,t,u,v=1}^{d} x_{(s,t)(u,v)}|v\rangle\langle u|s\rangle\langle t|& =\sum_{s,t,u,v=1}^{d}\delta_{us}x_{(s,t)(u,v)}|v\rangle\langle t|\\
& =\sum_{s,t,v=1}^{d}x_{(s,t)(s,v)}|v\rangle\langle t|\\
& =I_{d}=\sum_{t,v=1}^d\delta_{tv}|v\rangle\langle t|.\\
\end{aligned}
\end{equation*}
Therefore, we must have $$\sum_{s=1}^dx_{(s,t)(s,v)}=\delta_{tv}=\sum_{s=1}^dx_{((s-1)d+t)((s-1)d+v)}$$ for $t,v=1,2,...,d$, which is just $\text{Tr}_1X=I_d$.
\end{IEEEproof}

\section{PROOF OF THEOREM \ref{binvert}}\label{app2}
\begin{IEEEproof}
Using equation (\ref{property2}), we vectorize equation (\ref{betadef}) to obtain
\begin{equation}\label{app1}
(E_k^*\otimes E_j)\text{vec}(\rho_m)=\sum_n \beta_{mn}^{jk}\text{vec}(\rho_n).
\end{equation}
Let $\{W(j,k)\}_{j,k=1}^{d^2}$ be a family of $d^4$ matrices. The matrix $\{W(j,k)\}$ is $d^2\times d^2$ and its element in position $(m,n)$ is the number $\beta_{nm}^{jk}$. Let $V=(\text{vec}(\rho_1),\text{vec}(\rho_2),...,\text{vec}(\rho_{d^2}))$. From equation (\ref{app1}) we have
\begin{equation}
(E_k^*\otimes E_j)V=VW(j,k).
\end{equation}

Since $\{\rho_m\}_{m=1}^{d^2}$ is a set of linearly independent matrices forming a basis of the space $\mathbb C_{d\times d}$, $V$ must be invertible. Therefore we know
\begin{equation}\label{wexpress1}
W(j,k)=V^{-1}(E_k^*\otimes E_j)V.
\end{equation}

$Sufficiency:$ Since $\{E_i\}_{i=1}^{d^2}$ is a set of linearly independent matrices forming a basis of the space $\mathbb C_{d\times d}$, we know $\{E_k^*\otimes E_j\}_{j,k=1}^{d^2}$ is a set of linearly independent matrices forming a basis of the space $\mathbb C_{d^2\times d^2}$. Therefore, $\{W(j,k)\}_{j,k=1}^{d^2}$ is also a set of linearly independent matrices forming a basis of $\mathbb C_{d^2\times d^2}$. We then know $\{\text{vec}(W(j,k)^T)\}_{j,k=1}^{d^2}$ is a set of linearly independent column vectors forming a basis of the space $\mathbb C_{d^4\times 1}$, which leads to the conclusion that
\begin{equation}\label{bexpress1}
\begin{array}{rl}
B=&[\text{vec}(W(1,1)^T),\text{vec}(W(1,2)^T),...,\text{vec}(W(2,1)^T),\\
&\ \ \ \ \ \ \ \ \text{vec}(W(2,2)^T),...,\text{vec}(W(d^2,d^2)^T)]\\
\end{array}
\end{equation}
must be invertible.

$Necessity:$ When $B$ is invertible, from equation (\ref{bexpress1}) we know that $\{\text{vec}(W(j,k)^T)\}_{j,k=1}^{d^2}$ is a set of linearly independent column vectors forming a basis of the space $\mathbb C_{d^4\times 1}$. Therefore $\{W(j,k)\}_{j,k=1}^{d^2}$ is a set of linearly independent basis of $\mathbb C_{d^2\times d^2}$, and from equation (\ref{wexpress1}) $\{E_k^*\otimes E_j\}_{j,k=1}^{d^2}$ is also a set of linearly independent basis of $\mathbb C_{d^2\times d^2}$.

Now suppose that $\{E_i\}_{i=1}^{d^2}$ is not linearly independent. Then from equation (\ref{wexpress1}), one can easily prove $\{E_k^*\otimes E_j\}_{j,k=1}^{d^2}$ is not linearly independent, which leads to a contradiction. Hence, we have proved necessity.
\end{IEEEproof}

\section{PROOF OF THEOREM \ref{proequiv}}\label{app3}
\begin{IEEEproof}
We follow the notations in the Proof of Theorem \ref{binvert}. Since $\{\rho_m\}_{m=1}^{d^2}$ is a set of normal orthogonal basis of space $\mathbb C_{d\times d}$, we know $V$ is unitary. Therefore, we know
\begin{equation}\label{wexpress2}
W(j,k)=V^{\dagger}(E_k^*\otimes E_j)V.
\end{equation}

$Sufficiency:$ Since $\{E_i\}_{i=1}^{d^2}$ is a set of normal orthogonal basis of the space $\mathbb C_{d\times d}$, we have $$\delta_{(p,q)(k,j)} =\delta_{pk}\delta_{qj}=\langle E_p^*,E_k^*\rangle\langle E_q, E_j\rangle=\langle E_p^*\otimes E_q, E_k^*\otimes E_j\rangle,$$ which means $\{E_k^*\otimes E_j\}_{j,k=1}^{d^2}$ is a set of normal orthogonal basis of the space $\mathbb C_{d^2\times d^2}$. Therefore from (\ref{wexpress2}), we know that $\{W(j,k)\}_{j,k=1}^{d^2}$ is also a set of normal orthogonal basis of the space $\mathbb C_{d^2\times d^2}$. Hence $B$ must be unitary.

$Necessity:$ Since $B$ is unitary, from (\ref{bexpress1}) we know $\{W(j,k)\}_{j,k=1}^{d^2}$ is a set of normal orthogonal basis of $\mathbb C_{d^2\times d^2}$. According to (\ref{wexpress2}), we know $\{E_k^*\otimes E_j\}_{j,k=1}^{d^2}$ is also a set of normal orthogonal basis of $\mathbb C_{d^2\times d^2}$. Hence, we have
\begin{equation}\label{eqq1}
\langle E_p^*\otimes E_q,E_k^*\otimes E_j\rangle=\delta_{(p,q)(k,j)}=\delta_{pk}\delta_{qj}=\langle E_p^*,E_k^*\rangle\langle E_q, E_j\rangle.
\end{equation}

Now we concentrate on the third equality in (\ref{eqq1}). Setting $p=k=q=j$, we obtain $1=|\langle E_j, E_j\rangle|^2$. Since $\langle E_j, E_j\rangle=\text{Tr}(E_j^\dagger E_j)$ is a positive real number, we must have $\langle E_j, E_j\rangle=1$ for every $j=1,2,...,d^2$. Setting $p=k$, we obtain $\delta_{qj}=\langle E_q, E_j\rangle$, which means that $\{E_i\}_{i=1}^{d^2}$ is a set of normal orthogonal basis of the space $\mathbb C_{d\times d}$.
\end{IEEEproof}

\section{EXAMPLES OF ELIMINATING MULTIVALUED SOLUTIONS}\label{appendix1}
\begin{example}\label{example2}
Let $||\cdot||_x$ be any submultiplicative matrix norm (i.e., $||\cdot||_x$ satisfies (\ref{propertyhs2})). Suppose we know \emph{a priori} that $||H||_x$ is upper bounded by a known value $h_m$. Then we can set the evolution time $t<\frac{\pi}{2h_m}$.

The proof is straightforward. \emph{Theorem 1} in Chapter 10.3 of \cite{lancaster 1985} states that the absolute value of the eigenvalue of any matrix is no larger than the submultiplicative norm of the matrix. Hence, the prior knowledge in Example \ref{example2} is a sufficient condition for Assumption \ref{assump} to be satisfied.

\end{example}

\section{PROOF OF LEMMA \ref{lemma1}}\label{app4}
\begin{IEEEproof}
\begin{equation}
\begin{array}{ll}
||e^{i\theta}b-c||^2& =(e^{-i\theta}b^\dagger-c^\dagger)(e^{i\theta}b-c)\\
& =b^\dagger b+c^\dagger c -(e^{i\theta}bc^\dagger+e^{-i\theta}b^\dagger c).
\end{array}
\end{equation}
Let $b^\dagger c=re^{i\phi}$, where $r>0,\phi\in\mathbb R$. Then
\begin{equation}
\begin{array}{ll}||e^{i\theta}b-c||^2& =b^\dagger b+c^\dagger c-(re^{i\theta}e^{-i\phi}+re^{-i\theta}e^{i\phi})\\
& = b^\dagger b+c^\dagger c-2r\cos(\theta-\phi).
\end{array}
\end{equation}
Therefore, we should take $\theta=\phi$ to obtain
\begin{equation}
\min_\theta||e^{i\theta}b-c||^2=b^\dagger b+c^\dagger c-2r.
\end{equation}
We have
\begin{equation}
\begin{array}{ll}
||bb^\dagger-cc^\dagger||^2& =\text{Tr}(bb^\dagger bb^\dagger+cc^\dagger cc^\dagger-2bb^\dagger cc^\dagger)\\
& =(b^\dagger b)^2+(c^\dagger c)^2-2r^2.
\end{array}
\end{equation}

From the Cauchy-Schwartz inequality, $$r=|\langle b,c\rangle|\leq||b||\cdot||c||=\sqrt{b^\dagger b}\sqrt{c^\dagger c}.$$ We thus have
\begin{equation}
\begin{array}{rl}
&\frac{b^\dagger b+c^\dagger c}{2}\min_\theta||e^{i\theta}b-c||^2\\
=&\frac{1}{2}(b^\dagger b+c^\dagger c)^2-r(b^\dagger b+c^\dagger c)\\
\leq& (b^\dagger b)^2+(c^\dagger c)^2-r(b^\dagger b+c^\dagger c)\\
\leq&(b^\dagger b)^2+(c^\dagger c)^2- 2r\sqrt{b^\dagger b c^\dagger c}\\
\leq&(b^\dagger b)^2+(c^\dagger c)^2-2r^2\\
=& ||bb^\dagger-cc^\dagger||^2.\\
\end{array}
\end{equation}

On the other hand,
\begin{equation}
\begin{array}{rl}
&(\sqrt{b^\dagger b}+\sqrt{c^\dagger c})^2\min_\theta||e^{i\theta}b-c||^2\\
=&(b^\dagger b+c^\dagger c+2\sqrt{b^\dagger b}\sqrt{c^\dagger c})(b^\dagger b+c^\dagger c-2r)\\
=&(b^\dagger b+c^\dagger c)^2-4r\sqrt{b^\dagger b}\sqrt{c^\dagger c} \\
& +2(b^\dagger b+c^\dagger c)(\sqrt{b^\dagger b}\sqrt{c^\dagger c}-r)\\
\geq&(b^\dagger b)^2+(c^\dagger c)^2+2b^\dagger bc^\dagger c-4r\sqrt{b^\dagger b}\sqrt{c^\dagger c}\\
=&(b^\dagger b)^2+(c^\dagger c)^2+2(\sqrt{b^\dagger b}\sqrt{c^\dagger c}-r)^2-2r^2\\
\geq& (b^\dagger b)^2+(c^\dagger c)^2-2r^2\\
=& ||bb^\dagger-cc^\dagger||^2.
\end{array}
\end{equation}
\end{IEEEproof}


\section*{ACKNOWLEDGEMENT}

The authors would like to thank Zhibo Hou and Guo-Yong Xiang for helpful discussions.

\ifCLASSOPTIONcaptionsoff
  \newpage
\fi

\end{document}